\documentstyle[pra,aps,multicol,psfig]{revtex}

\begin{document}


\draft

\title{Elastic and inelastic collisions of $^6$Li atoms in magnetic and optical
traps}

\author{M. Houbiers,$^1$ H. T. C. Stoof,$^1$ W. I. McAlexander,$^2$
and R. G. Hulet$^2$} 
\address{$^1$Institute for Theoretical Physics, University of Utrecht,
         Princetonplein 5, 3584 CC Utrecht, The Netherlands \\
         $^2$Physics Department and Rice Quantum Institute,
          Rice University, Houston, Texas 77251-1892}
        
\maketitle

\begin{abstract}
We use a full coupled channels method to calculate collisional
properties of magnetically or optically trapped ultracold $^6$Li.
The magnetic field dependence of the $s$-wave scattering lengths
of several mixtures of hyperfine states are determined, as are the
decay rates due to exchange collisions.
In one case, we find Feshbach resonances at $B=0.08$~T and $B=1.98$~T.
We show that the exact coupled channels calculation is well 
approximated over the entire range of magnetic fields by a simple 
analytical calculation.
\end{abstract}


\begin{multicols}{2}
The observation of Bose-Einstein condensation in atomic alkali gases 
\cite{JILA,rice,MIT} has triggered an enormous interest in 
degenerate atomic gases. At present, one of the most important
goals is to achieve quantum degeneracy in a fermionic gas. 
In the case of fermionic $^6$Li, it has been shown theoretically 
that a BCS transition to a superfluid state could be realized 
at a critical temperature on the order of temperatures obtained in
the BEC experiments \cite{us}. 
This relatively high critical temperature is due to the fact that
$^6$Li has a very large and negative triplet $s$-wave scattering length
$a_T=-2160 a_0$ \cite{randy2}, and that at sufficiently large magnetic 
fields a mixture of the upper two hyperfine states $|6\rangle$
and $|5\rangle$ is essentially electron-spin polarized \cite{hyperfine}. 

The disadvantage of such a large triplet $s$-wave scattering length
is that the exchange and dipolar relaxation rates for the gas
are also anomalously large. Nevertheless, to suppress this decay, 
one can apply a magnetic bias field. 
In Ref.~\cite{us}, we used the distorted wave Born approximation (DWBA) 
to calculate the corresponding decay rate constants for these decay 
processes and found that at large magnetic fields $B>10$ T, 
the dipolar rates are dominant, but at smaller magnetic fields, the 
exchange rates greatly exceed those due to the dipolar interaction. 
However, the DWBA is only valid at magnetic fields $B > 0.1$ T, and
we were at that time unable to make predictions at lower, experimentally
more convienent fields.
 
The aim of the present paper is to provide useful information
on the $s$-wave scattering length and exchange decay rate constants
at lower magnetic fields. 
In view of the ongoing experiments, we will concentrate on 
collisions involving the
following antisymmetrized hyperfine states: $|\{65\}\rangle$, 
$|\{64\}\rangle$, $|\{54\}\rangle$, and $|\{21\}\rangle$. The first three 
mixtures contain states that are low-field seeking at sufficiently high field,
and therefore, can be confined in a magnetic trap.
In contrast, the combination $|\{21\}\rangle$ cannot be magnetically trapped,
but can be confined in a far off-resonance optical trap. 
We do not consider any other combination of high-field seeking states, 
since the $|\{21\}\rangle$ mixture cannot decay through collisions
and is therefore most favorable experimentally. 
In addition, Van Abeelen {\it et al.} \cite {verhaar} have 
already considered the $|\{62\}\rangle$ combination, 
which is low-field seeking at very weak magnetic fields 
$B\leq 26 \times 10^{-4}$~T, but does not have 
as large an $s$-wave scattering length as some other combinations \cite{ian}. 

To obtain the magnetic field dependence of the $s$-wave scattering length 
and exchange decay rate constants in the various cases, we perform a full
coupled channels (CC) calculation \cite{stoof},
using the most up-to-date singlet
and triplet potentials $V_0(r)$ and $V_1(r)$ \cite{randy2}. 
Furthermore, we show that the rate constants and the scattering lengths
found using the CC calculation can be obtained analytically using a simple 
approximation that we call the {\it asymptotic} boundary condition (ABC)  
approximation \cite{feshbach}.  

If the thermal energy is much smaller than the hyperfine plus Zeeman energy, 
$\Delta_{\alpha\beta \rightarrow \alpha'\beta'}$,
gained in the transition from the incoming state $|\{\alpha \beta\}\rangle$ 
to an outgoing state $|\{\alpha' \beta' \}\rangle$, the corresponding exchange
rate constant is given by the zero temperature expression \cite{stoof}
\begin{eqnarray}
G_{\alpha\beta\rightarrow\alpha'\beta'} & = &
\lim_{k_{\alpha\beta} \rightarrow 0} \frac{\pi \hbar}{\mu k_{\alpha\beta}} 
\left| S_{\{\alpha' \beta'\} 00, \{\alpha\beta\} 00}(k_{\alpha\beta}) ... 
\right. \nonumber \\ 
\label{rates}
& & \hspace{3cm} \left. ... - 
\delta_{\{\alpha' \beta'\},\{\alpha\beta\}} \right|^2.  
\end{eqnarray}
In this expression, $\mu$ is the reduced mass of the two $^6$Li atoms, 
$\hbar k_{\alpha\beta}$\ is the relative momentum of the 
incoming particles in state $|\{\alpha\beta\}\rangle$, and the matrix 
$S$ is the scattering matrix of the multi-channel problem with
angular momentum quantum numbers $l=0$ and $m=0$. The number of channels 
being coupled is determined by the fact that the central interaction
$V^c(r) = V_0(r) {\cal P}^0 + V_1(r) {\cal P}^1$ 
cannot change the total nuclear plus electron-spin projection 
of the two-particle wave function along the magnetic field. So for example, 
the state $|\{65\}\rangle$ can only decay to $|\{61\}\rangle$.
While the nondiagonal part of the unitary S-matrix gives the decay rates,
the diagonal part of the S-matrix gives the $s$-wave scattering length via  
\begin{equation}
S_{\{\alpha\beta\},\{\alpha\beta\}} \rightarrow \exp{(- 2i k_{\alpha\beta}
a_{\alpha\beta} )} \simeq 1 - 2i k_{\alpha\beta} a_{\alpha\beta}\ ,
\label{aswave}
\end{equation} 
when $k_{\alpha\beta} \rightarrow 0$. The goal of the CC 
calculation is to determine the S-matrix, but rather
than giving a detailed description of the calculation, we refer to
Ref.~\cite{stoof} and merely present the results here.

\begin{figure}[htbp]
\psfig{figure=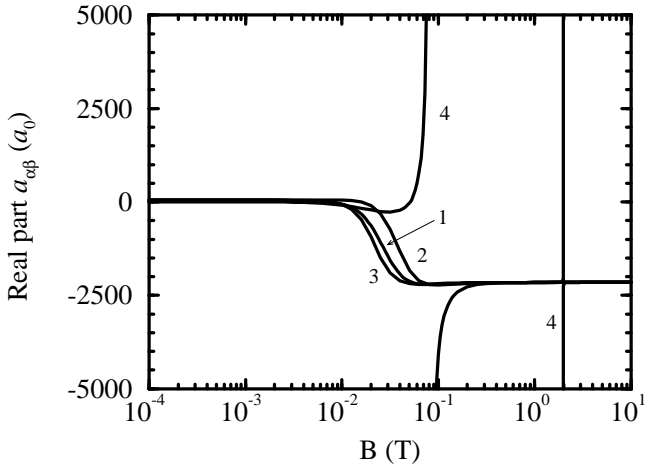}
\caption{\protect\narrowtext
Real part of zero-temperature $s$-wave scattering length 
for 1) 65, 2) 54, 3) 64, and 4) 21 collisions as a function of the 
magnetic field.}
\label{fig1}
\end{figure}

In Fig.~\ref{fig1}, the real part of the $s$-wave scattering length 
is plotted for the four cases of interest. 
Note that, in the first three cases, the $s$-wave scattering lengths
attain the large and negative value of $a_T = -2160 a_0$, only for 
$B \geq 0.05$~T.
At lower magnetic fields, the scattering lengths become $47 a_0$ at zero 
magnetic field for the first three cases, and zero in the latter case. 
This effect is explained in detail below.
The scattering length $a_{21}$ exhibits Feshbach resonances \cite{tiesinga}
at $B \simeq 0.08$ T and $B \simeq 1.98$ T. Physically, a Feshbach resonance  
arises whenever the Zeeman energy of the incoming 
wave function $|\{21\}\rangle$, which is almost purely triplet,
coincides with a bound-state energy of the singlet potential 
$V_0(r)$. Only the state $|\{21\}\rangle$ can exhibit Feshbach resonances 
at the magnetic fields of interest since it is the only combination that 
has a negative total energy (about $-2 \mu_e B$). 
The precise location of these resonances is shifted slightly due
to the hyperfine coupling, but their positions can be accurately computed since
the binding energies of the relevant singlet states, $v=38$ and $v=37$,
are accurately calculated from the experimentally determined 
singlet potential \cite{randy2}. 

\begin{figure}[htbp]
\psfig{figure=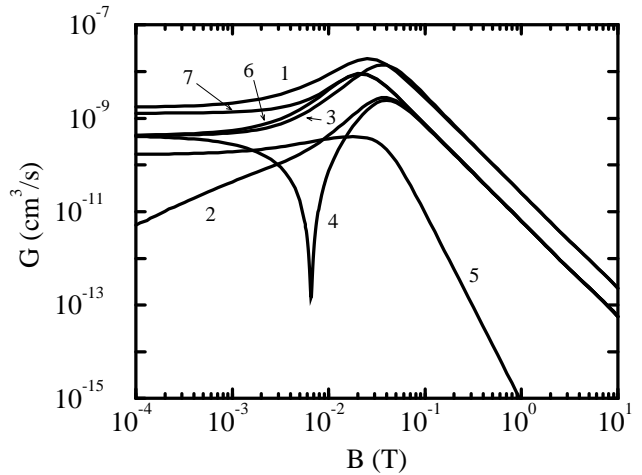}
\caption{\protect\narrowtext
Exchange decay rate constants for the processes 
1) $65 \rightarrow 61$, 2) $54 \rightarrow 63$, 3) $54 \rightarrow 52$,
4) $54 \rightarrow 41$, 5) $54 \rightarrow 21$,
6) $64 \rightarrow 62$, 7) $64 \rightarrow 51$
as a function of the magnetic field strength.}
\label{fig2}
\end{figure}

The decay rate constants resulting from the CC 
calculation are, for all possible exchange processes, 
plotted in Fig.~\ref{fig2}. The sharp minimum in the rate 
$G_{54\rightarrow 41}$ is a result of the fact that
the matrix element $\langle \{41\}| V^c | \{54\}\rangle$
is zero if $\tan \theta_- = 1/\sqrt{2}$. However, the state 
$|\{54\}\rangle$ still remains coupled to $|\{41\}\rangle$ through
the other elements in the ($5\times5$) coupling matrix. As a result
of this interference, the rate constant is never zero, and the
location of the minimum is 
slightly shifted from the magnetic field at which 
$\tan \theta_- = 1/\sqrt{2}$. 

The second remark regarding Fig.~\ref{fig2} concerns the behavior 
of the rate constants at large $B$-fields where $\sin \theta_{\pm}
\simeq \theta_{\pm} \propto 1/B$. The slope of each curve is determined 
by the magnetic field dependence of $\sqrt{B} \left| \langle 
\{\alpha' \beta'\}|V^c| \{\alpha\beta\}\rangle \right|^2$. After some
algebra, one can show that for curves 2, 4, 6 and 7, the relevant matrix 
element is equal to $\theta_{\pm} (V_1-V_0)/2 \equiv \theta_{\pm}
V^{ex}/2$, which is a factor $1/2$ smaller than 
in the cases 1 and 3. Therefore, the decay rates are a factor of 4 smaller 
than in the latter cases, but the slopes are identical. On the other hand,
the matrix element $\langle \{21\}|V^c|\{54\}\rangle = 
3\theta_+\theta_- V^{ex}/2$, has an additional factor of $1/B$, and thus the 
slope of curve 5 is a factor of $7/3$ smaller than the others.

A comparison of the rate constants shows that the
$|\{64\}\rangle$ combination is slightly more favorable than the 
$|\{65\}\rangle$ combination. 
The total rate constant of the former combination 
is only half the rate constant of the latter at large magnetic fields.
At $B=0.1$ T, they amount to $1.49\times10^{-9}$ cm$^3$/s and
$2.74\times 10^{-9}$ cm$^3$/s, respectively.

We now show that the CC results can be accurately reproduced using 
a simple analytical calculation. We consider here only the 
65 case, but calculation of the other combinations is analogous,
and gives similar accuracy. 
In the ABC approximation, space is divided in two regions. In the
interior region $r < R$, the singlet and triplet potentials dominate,
while in the outer region $r>R$, the hyperfine energies prevail. The boundary
$R$ is chosen between the points where the exchange potential
$V^{ex}(r)$ is of the order of the hyperfine constant 
$a_{hf}$, and the point where the triplet and singlet potentials
themselves are of order $a_{hf}$. In our case this is between $r=28 a_0$ and
$r=62 a_0$, and below we will use as examples $R=40 a_0$ and $R=60 a_0$.

In the interior region the hyperfine splitting is neglected, so that the  
channel wave functions are just linear combinations of the singlet and
triplet scattering wave functions, which we approximate by their
asymptotic forms $A_S\sin {k_{\alpha\beta}(r-a_S)}$ and  
$A_T\sin {k_{\alpha\beta}(r-a_T)}$, in the limit 
of small momentum $k_{\alpha\beta}$. The known values of the 
scattering lengths are $a_S = + 45.5 a_0$ and $a_T=-2160 a_0$, 
respectively \cite{randy2}. In the exterior region, we neglect the
central interaction $V^c$, so the channel wave function is a
plane wave $\exp{(-ik_{\alpha\beta}r)}$ for the incoming channel
$|\{\alpha \beta\}\rangle$, and $\tilde{S}_{\{\alpha'\beta'\},
\{\alpha\beta\}} \exp{(+ik_{\alpha'\beta'}r)}$
for all outgoing channels $|\{\alpha'\beta'\}\rangle$. 
The wave numbers $k_{\alpha'\beta'}$ depend on the energy gained by the
transition to the respective outgoing channel, that is, 
$k_{\alpha'\beta'} =\sqrt{k^2_{\alpha\beta} + 2\mu
\Delta_{\alpha\beta\rightarrow\alpha'\beta'} /\hbar^2}$. 

\begin{figure}[htbp]
\psfig{figure=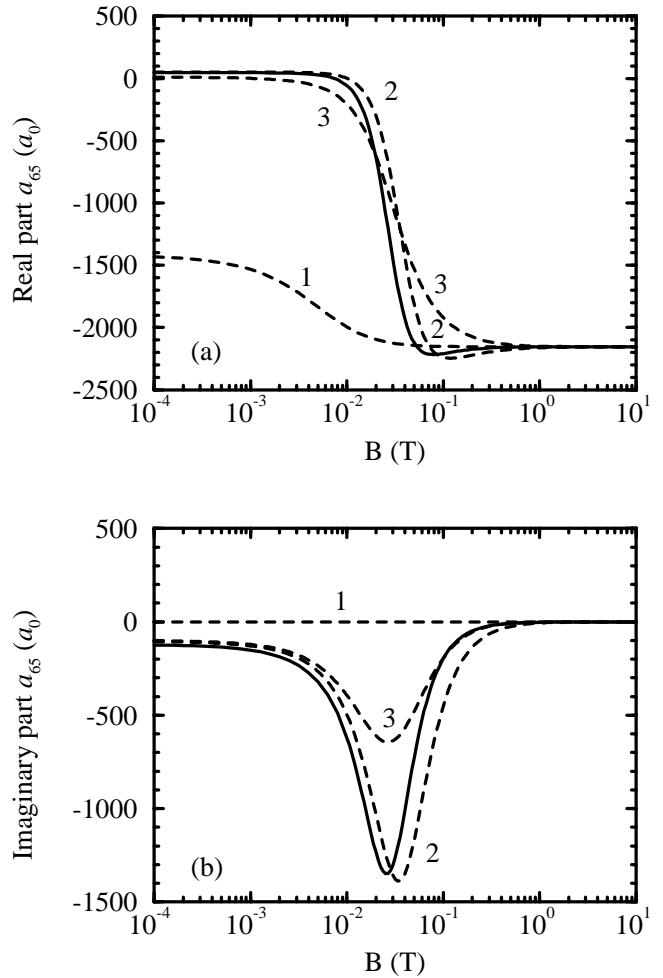}
\caption{\protect\narrowtext
The $a)$ real and $b)$ imaginary part of the $s$-wave scattering 
length $a_{65}$. The solid line is the CC 
value, which goes to the triplet scattering length for large
magnetic fields. The dashed lines are $1)$ the DIS approximation, 
$2)$ the ABC approximation with $R=40a_0$, and $3)$ the ABC  
approximation with $R=60 a_0$.}   
\label{fig3}
\end{figure}

The yet unknown amplitudes $A_S$, $A_T$, and $\tilde{S}_{\{\alpha'
\beta'\},\{\alpha\beta\}}$ must be determined by imposing 
continuity and differentiability to the wave function in each 
channel at $r=R$. To do so, we need to find the precise
linear combination of singlet and triplet wave function in the
interior region. 
In terms of the basis $|S\:M_S;I\:M_I\rangle$, where 
${\bf S=s_1+s_2}$ and ${\bf I=i_1 +i_2}$, are the total electron and 
nuclear spin of the two-atom system respectively, we find 
\begin{eqnarray*}
|\{65\}\rangle & = & \sin \theta_+ |0\:0; 2\:2\rangle + 
\cos \theta_+ |1\:1;1\:1\rangle, \\ 
|\{61\}\rangle & = & \cos \theta_+ |0\:0; 2\:2\rangle - 
\sin \theta_+ |1\:1;1\:1\rangle.
\end{eqnarray*} 
\noindent 
Evidently, for the channel wave functions 
we have to take linear combinations with the same coefficients, so, 
for example the
wave function of the $|\{65\}\rangle$ channel in the interior region becomes
$\psi_{65}(r) = A_S \sin{\theta_+} \sin{k_{\alpha \beta}(r-a_S)} 
+ A_T \cos{\theta_+} \sin{k_{\alpha \beta}(r-a_T)}$. 
It can be shown easily that the coefficients 
$\tilde{S}_{\{\alpha'\beta'\},\{\alpha\beta\}}$ are related with 
the S-matrix \cite{stoof} according to
\begin{equation}
S_{\{\alpha'\beta'\},\{\alpha\beta\}} = - \sqrt{\frac{k_{
\alpha'\beta'}}{k_{\alpha\beta}} } 
\tilde{S}_{\{\alpha'\beta'\},\{\alpha\beta\}}. 
\label{smatrix}
\end{equation}
For the case of the incoming $|\{65\}\rangle$-state, we are only interested 
in the constants $\tilde{S}_{\{65\},\{65\}}$ and $\tilde{S}_{\{61\},\{65\}}$,
and therefore can eliminate the constants $A_S$ and $A_T$ from the
$4\times4$ set of complex equations. The resulting $2\times2$ set of complex
linear equations for these amplitudes can be solved in orders of $k_{65}$, 
and we find to first order in $k_{65}$ 
\[     
\tilde{S}_{\{65\},\{65\}} =  -1 + 2i k_{65} \left\{ R + 
\frac{ik^{(0)}_{61}(R-a_S)(R-a_T)}{D(k^{(0)}_{61})} \right. ... 
\]
\begin{mathletters}
\begin{equation}
... - \left.
\frac{\cos^2\theta_+(R-a_T) + 
\sin^2\theta_+ (R-a_S)}{D(k^{(0)}_{61})} \right\},
\label{s1}
\end{equation}
\begin{equation}
\tilde{S}_{\{61\},\{65\}} =  \frac{2 i k_{65} \cos \theta_+ \sin \theta_+
(a_S - a_T)}{ D(k^{(0)}_{61})},
\label{s2}
\end{equation}
\label{s}
\end{mathletters}

\noindent
where 
$k^{(0)}_{61} = \sqrt{2\mu \Delta_{65 \rightarrow 61} /\hbar^2}$, and
$D(k^{(0)}_{61})= 1 - i k^{(0)}_{61} [ \cos^2\theta_+(R-a _S)
 + \sin^2\theta_+ (R-a_T) ]$.
Combining Eqs.~(\ref{s1}), 
(\ref{smatrix}) and (\ref{aswave}) we find the zero-momentum 
$s$-wave scattering length $a_{65}$, which is plotted in
Fig.~\ref{fig3}, together with the CC curve, and the result of a 
degenerate internal states (DIS) approximation \cite{stoof}. 
Fig.~\ref{fig3}$a$ shows the real part, while Fig.~\ref{fig3}$b$ gives the 
imaginary part.

The ABC approximation is clearly much better than the DIS approximation. 
To obtain the DIS result, the hyperfine level splitting is negelected, 
which is equivalent to setting $k^{(0)}_{61} = 0$ in Eq.~(\ref{s1}), 
resulting in Re$(a_{65}) =
a_T \cos^2\theta_+  + a_S \sin^2 \theta_+$ and Im$(a_{65}) =0$.
For hydrogen and deuterium, terms containing $k^{(0)}_{\alpha'\beta'} 
(R-a_{T,S})$ can indeed be neglected compared to $1$ at lower
magnetic fields, so that DIS is a rather good approximation in that case.
However, for $^6$Li with its anomalously large value of $a_T$, this is not
the case, and one must then take terms of order $k^{(0)}_{61}(R-a_T)$ 
into account. Also, for heavier alkalies, the DIS approximation becomes 
increasingly bad even
at lower magnetic fields, due to the dependence of $k^{(0)}_{\alpha'\beta'}$
on the atomic mass and hyperfine constant. 

It is straightforward to show that at large magnetic fields, the ABC
approximation gives Re$(a_{65}) = a_T$, and Im$(a_{65})=0$, independent
of the choice of $R$. At $B=0$, we find Re$(a_{65}) = 3a_S -2 R \simeq
57 a_0$ for $R=40 a_0$, and 
Im$(a_{65}) = - 1/(k^{(0)}_{61} \tan^2\theta_+) \simeq -100 a_0$,
in surprisingly good agreement with the results of the exact CC calculation.
Note that the imaginary part becomes rather large and contributes
significantly to the elastic cross-section $4 \pi |a_{65}|^2$ at
lower magnetic field.

\begin{figure}[htbp]
\psfig{figure=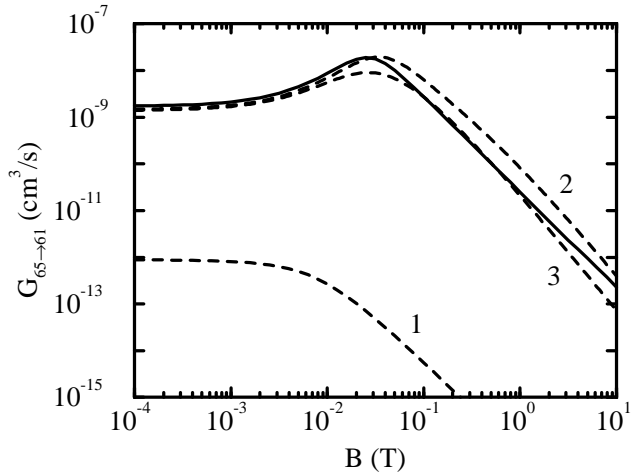}
\caption{\protect\narrowtext
Exchange decay rate $G_{65 \rightarrow 61}$ of the $^6$Li gas. 
The solid line gives the CC result. The dashed lines represent $1)$ the 
DIS approximation, $2)$ the ABC approximation with $R=40a_0$, and $3)$ the
ABC approximation with $R=60 a_0$.}   
\label{fig4}
\end{figure}
Similarly, one can find the decay rate constant $G_{65\rightarrow 61}$
from Eq.~(\ref{s2}) combined with Eqs.~(\ref{smatrix}) and (\ref{rates}).
The results are plotted in Fig.~\ref{fig4}.
Again, agreement with the exact CC calculation is very good, whereas the
DIS approximation is in error by several orders of magnitude.
We believe that application of the ABC approximation to other alkali 
atoms for which the potentials are not so well known, can also give an accurate 
estimate for the collisional properties of interest \cite{cc}. 
Of course, the
results will vary slightly with the actual choice of $R$. It is 
difficult to predict {\it a priori} which $R$ reproduces the exact 
results for an
arbitrary atom best, but any $R$ chosen as indicated previously,  
will give at least a good order of magnitude estimate. The
DIS approximation on the other hand, can be off by many orders of magnitude.
  
In summary, we have calculated the exchange decay rate constants and
$s$-wave scattering length for four different combinations of hyperfine 
states of $^6$Li by means of a coupled channels calculation. 
The scattering lengths attain their large and negative value of $-2160 a_0$
only for $B \geq 0.05$~T. We found that the $|\{21\}\rangle$ 
combination has Feshbach resonances at $B=0.08$~T and $B=1.98$~T.
In a magnetic trap, the lifetime of the $|\{64\}\rangle$ 
combination is slightly higher than the one for the $|\{65\}\rangle$
combination. The results of the exact calculation can be obtained 
relatively easily and accurately 
with the asymptotic boundary condition approximation.  

We acknowledge useful discussions with Jean Dalibard, Marc Mewes
and John Tjon.
The work at Rice was supported by the National Science Foundation,
NASA, the Texas Advanced Technology Program, and the Welch Foundation.

\end{multicols}






\begin{references}
\bibitem{ian} The $s$-wave scattering length of the $|\{62\}\rangle$ 
	combination is about $-1615 a_0$, for $B \leq 26 \times 10^{-4}$~T.
\end{references}
\end{document}